# CCC / Code 8.7
# Applying AI in the Fight Against Modern Slavery

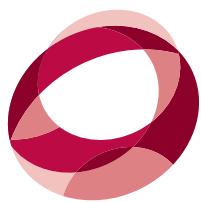

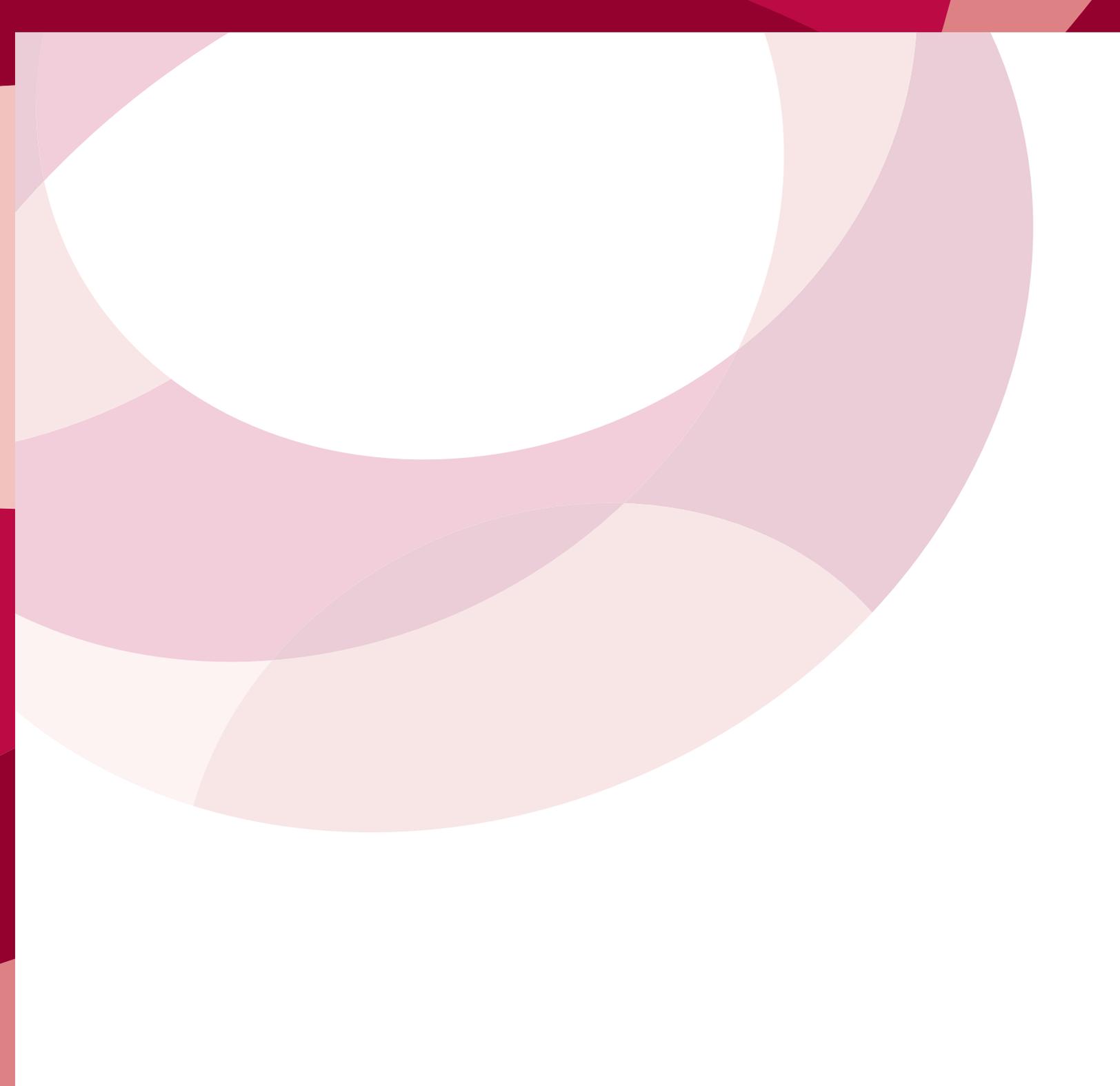

The material is based upon work supported by the National Science Foundation under Grant No. 1734706. Any opinions, findings, and conclusions or recommendations expressed in this material are those of the authors and do not necessarily reflect the views of the National Science Foundation.

# CCC / Code 8.7
# Applying AI in the Fight Against Modern Slavery

**Steering Committee:**

Nadya Bliss (Arizona State University)
Mark Briers (Turing Institute)
Alice Eckstein (UNU-CPR)
James Goulding (University of Nottingham)
Daniel Lopresti (Lehigh University)*
Anjali Mazumder (Turing Institute)
Gavin Smith (University of Nottingham)
*Lead Workshop Organizer

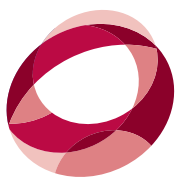

CCC
Computing Community Consortium
Catalyst

## Broad Overview

On any given day, tens of millions of people find themselves trapped in instances of modern slavery. The terms "human trafficking," "trafficking in persons," and "modern slavery" are sometimes used interchangeably to refer to both sex trafficking and forced labor. Human trafficking occurs when a trafficker compels someone to provide labor or services through the use of force, fraud, and/or coercion [HTSD2020].

The wide range of stakeholders in human trafficking presents major challenges. Direct stakeholders are law enforcement, NGOs and INGOs, businesses, and local/planning government authorities. Survivors additionally act as direct stakeholders: their lived experiences are vital to identifying and characterizing trafficking, and proposed solutions must be measured by their effectiveness in helping survivors. Indirect stakeholders include the public who benefit from reductions in criminal activity, programs and services (schools, family support programs, juvenile corrections) that operate in environments at high risk for sex trafficking, and increases in overall community wellness/safety. Viewed from a very high level, all stakeholders share in a rich network of interactions that produce and consume enormous amounts of information. The problems of making efficient use of such information for the purposes of fighting trafficking while at the same time adhering to community standards of privacy and ethics are formidable.

At the same time they help us, technologies that increase surveillance of populations can also undermine basic human rights. The IEEE Global Initiative on Ethics of Autonomous and Intelligent Systems [IEEE2020] recently put forward a number of Artificial Intelligence responses to consider in light of the unique challenges brought about by the COVID-19 pandemic; community health and crimes like trafficking share a number of important parallels when AI methods are brought to bear. For example, the huge amount of sensory data (locations, calls, and text messages) collected from mobile devices have the potential to advance significant AI technologies based on user data extracted from sensor readings. However, the careless use or inadvertent release of such data could also seriously threaten user privacy and even safety [PUC 2017]. It is important that any kind of data collection be treated sensitively, not only for the immediate privacy concerns of the participants who provide their personal information, but in consideration of how that data could affect the survivor population as a whole.

In early March 2020, the Computing Community Consortium (CCC), in collaboration with the Code 8.7 Initiative, brought together over fifty members of the computing research community along with anti-slavery practitioners and survivors to lay out a research roadmap. The primary goal was to explore ways in which long-range research in artificial intelligence (AI) could be applied to the fight against human trafficking. Building on the kickoff Code 8.7 conference held at the headquarters of the United Nations in February 2019 [Code8.7], the focus for this workshop was to link the ambitious goals outlined in the *A 20-Year Community Roadmap for Artificial Intelligence Research in the US* (AI Roadmap) to challenges vital in achieving the UN's Sustainable Development Goal Target 8.7, the elimination of modern slavery [AI19].

It is important to note that while most research roadmaps take a positive view of the scientific advances they are promoting, trafficking is also an adversarial use case, and consumer-grade technologies are available to perpetrators just as they are available to law enforcement and NGOs. Not all AI is used for good. In the case of modern slavery, the stakes are high since the criminal exploitation of advanced AI could lead to life-or-death situations. It is critical that researchers proceed with an awareness of these possibilities and do their best to keep dangerous AI tools out of the hands of traffickers. An AI "arms race" is inevitable, but we must win it.

In the time since the workshop took place in early March 2020, the world has been slammed by a global pandemic. Social unrest has also rocked a number of countries, including the US. The following report is the outcome of the March 2020 workshop combined with subsequent discussions and other relevant sources produced as a result of the events of the past year.

## Technical Background

Individuals who have been trafficked or who are vulnerable to exploitation face pressures and demands throughout every minute of their day. The numerous choices they make determine the path they will take and can mean the difference between teetering on the brink of further exploitation, or beginning the arduous climb back toward freedom. These choices are limited to the opportunities they know are available to them, among the overwhelming set of possibilities that may be present in a given environment. In crisis situations, time becomes even more critical, and there is even less time to seek out the best option from traditional means of support. Looking toward the future for solutions, we might imagine a day when an interactive, intelligent, ethical personal assistant that is aware, deeply perceptive, adaptable, and fully conversant could take advantage of many opportunities to intervene. Such a technology would support the day-to-day choices that make a difference for vulnerable individuals walking the line between exploitation and freedom all while protecting their privacy and enhancing quality of life. This is one example of the kinds of linkages we shall describe between the AI Roadmap [AI19] and Code 8.7 goals.

Researchers know there are discrete populations vulnerable to exploitation. There are already big datasets available to understand environmental factors in vulnerability, and more are waiting to be harvested. Additionally, the lived experience of survivors adds important nuance to understand how these factors combine with contact with traffickers to create "tipping points" that channel vulnerable individuals toward situations where they will be exploited.

It is not just existing social services, law enforcement, and "dark web" datasets that can be mined for valuable insights, however. Moving beyond these traditional sources of data, the myriad of daily interactions representing the complete range of online and connected activities of an individual, group, or community are also incredibly valuable in formulating solutions. A deep awareness of what is going on in someone's life leading up to a crisis point is important when formulating a plan to try to help them.

How can AI be useful here? One approach might be to offer survivors opportunities to share data and expertise through interactions with machine learning systems, with the goal of creating interactive and flexible models to guide the intervention and disruption of trafficking patterns across a variety of contexts.

## Drawing from the AI Roadmap

In this report, we draw from the following ideas presented in the recent 20-Year Community Roadmap for AI Research [AI19]:

A.  Societal Drivers (Section 3.2.2): Societal drivers for AI systems lead to the creation of new tools that enhance health and quality of life.

B.  Meaningful Interaction with AI Systems (Section 3.2.3): Meaningful interaction with AI systems includes modeling and communicating mental states, alignment with human values and social norms, making AI systems trustworthy, preventing undesirable manipulation, and supporting interactions between people.

C.  Self-Aware Learning (Section 3.3.3): Self-aware learning refers to explanation and interpretability, durable machine learning systems, trustworthy learning and data provenance, and quantification of uncertainty.

D.  Contextualized AI (Section 3.1.4): Under contextualized AI we have social cognition; if AI is to serve as an effective collaborator, it must understand the interactions between individuals and their worldviews.

E.  The Science of Integrated Intelligence (Section 3.1.3): The goal of Integrated Intelligence is for computers to synthesize and integrate from a broad range of heterogeneous life-long experiences, like what humans can easily do.

F.  Knowledge Repositories (Section 3.1.5) and Open AI Platforms (Section 5.1.1): Knowledge repositories and open AI platforms include the capture and dissemination of knowledge, managing heterogeneous knowledge, and the creation of open, AI-ready data repositories.

G.  AI Ethics and Policy (Section 5.2.4.1.1): AI ethics and policy the incorporation of ethics and related responsibility principles as central elements in the design and operation of AI systems.

H.  Life-long and Life-wide Personal Assistant Platforms (Section 5.1.1.2.2): Life-long and life-wide personal assistant platforms (LPAs) will learn and adapt over their users' lifetime. They will detect situations for opportunistic training and education based on modeling the user and use this information to help an individual develop better, more complete and fruitful understandings of the world they live in.

It should be intuitively clear why each one of these capabilities will be important for AI to help achieve the goal of ending modern slavery. All of these constitute active areas of interest as identified by the AI research community.

## Vignettes

We now present a series of five vignettes to illustrate the various dimensions of this problem: the challenges and opportunities presented in applying AI to the fight against human trafficking. It should be noted that there is a large amount of "low hanging" fruit where technology can play a supporting role, and a sense of tremendous urgency because of the suffering that is taking place even as we write these words. While acknowledging the important work going on right now, our focus is on research objectives that will take years or even decades to achieve. This is the nature of computing and AI research, and it does not diminish our enthusiasm and support for colleagues working to address the crisis today.

Before each vignette, we give a short commentary on the key points that connect it to a subset of AI research topics. Then, in the next section, we provide a more detailed linking to the AI Roadmap [AI19], cutting across all of the vignettes.

### Vignette #1 (Perceptive Agents to Provide Survivor Support)

This first vignette is drawn from the perspective of a survivor. It is written with a specific focus on the "Meaningful Interaction," "Contextualized AI," "Life-Long and Life-Wide Personal Assistant Platforms," "Societal Drivers for Meaningful Interaction with AI Systems," and "Integrated Intelligence" topics. Look for the LPA interactions with Susan and see how the AI system learns about her and carefully guides her toward safe choices and provides the right level of assistance at the right time.

Vignette #1 (adapted from Vignette #13 in the AI Roadmap [AI19])

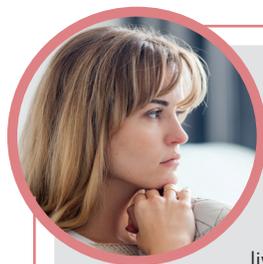

**Vignette 1**

Susan comes home to an eviction notice on her door, which gives her 24 hours to remove all of her possessions from her apartment. She has had problems with her landlord and feels that the eviction is unjustified and possibly illegal. But her immediate concern is to look for another place to live, so she asks her AI assistant to search for short-term housing in her city. The LPA returns a set of results, but in a side channel (visual or auditory) asks if something is wrong with her current apartment, since it knows that her lease runs for another eight months. She tells the system that she was evicted. Recognizing this as a potential tipping point leading toward exploitation of individuals in general terms, and knowing the details of Susan's past experiences having been trafficked in particular and her health, the AI system is careful to guide Susan toward options that are most likely to be safe and effective for her own personal circumstances.

**Vignette 1 Continued**

As one of its services, the AI system provides legal support. The system queries relevant tenant rights law for her city and state. The result is a set of legal documents, which are long and difficult for non-experts to read and understand. Realizing this, the LPA identifies critical parts of each document specifically relating to eviction in the context of Susan's situation. Susan can highlight parts that she does not understand, and ask for clarification; the AI system can respond either by generating explanations in language specifically targeted to Susan's level of understanding, which it knows well, because it knows her, or it can identify snippets of relevant documents for Susan to read herself. The AI system offers to connect her with nearby legal aid resources in the morning.

The LPA also provides Susan with a set of just-in-time social connections to individuals who have been in a similar situation and are willing to share their experiences, again taking special care to avoid individuals who might serve as a bridge between Susan and a situation where she would be exploited as she has in the past. This group can include individuals who have fought a similar eviction notice, individuals who have used short-term housing in the same area, others who have used free legal aid services, and other current and former tenants of the same landlord. Keeping in mind that these individuals may not have shared these experiences widely, the AI system needs to decide when and under what conditions to share what information to support Susan's interactions with the others. In addition to person-to-person connections, the LPA can aggregate social media content from these individuals related to their experiences, which will create a set of resources to help Susan deal with her situation both practically and emotionally.

While it is actively engaged in helping Susan, the AI system is also acquiring new knowledge that will be useful to researchers, social service workers, policy actors, and law enforcement professionals who are working to better understand and address vulnerabilities and tipping points in her neighborhood and more broadly across their jurisdictions.

## Vignette #2 (Perceptive Agents to Identify Tipping Points)

We now turn to a vignette from the perspective of an organization intervening where communities might become vulnerable to exploitation, or providing social services to victims and survivors. It is written with a specific focus on the "Self-Aware Learning," "Integrated Intelligence," and "Knowledge Repositories and Open AI Platforms." Note how the AI is able to mine rich data collected over long periods of time under widely varying circumstances to determine the factors that make someone a target for trafficking.

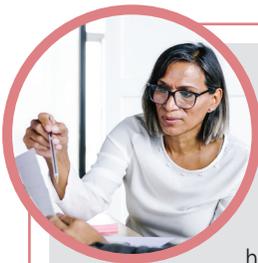

**Vignette 2**

Alex is a researcher working at an NGO to identify and disrupt the common behaviors of exploiters and trafficking recruiters in their community. They recognize that many factors, ranging from cultural norms to victim/target characteristics, drive the nature of the perpetrator's behaviors. Alex has funding to conduct research and has established preliminary relationships with a wide range of shelter and support organizations. In addition, Alex's work has the backing of key stakeholders, and they have been granted access to a wide range of valuable data collected in the form of knowledge repositories that are designed to be "AI-ready."

**Vignette 2 Continued**

Alex engineers their system using advanced AI to generate actionable insights in a survivor-informed manner. The AI is able to mine rich data collected over long periods of time and widely varying circumstances to determine the factors that make an individual or groups of individuals likely targets for trafficking, as well as the perpetrators who are seeking to take advantage of the situation. The full range of local, community, and regional effects are accounted for, including health, nutrition, education, employment, and politics. Mostly, however, the AI has learned from interacting with survivors and their lived experiences. The AI system will respond to complex queries phrased the way an expert might use when conferring with a colleague and not just simple "google-like" searches (e.g., to help law enforcement gather and process evidence for successful prosecutions), and also proactively notify the appropriate authorities when conditions seem to be heading down a path that would lead toward a new trafficking activity, proposing various ways in which this can be effectively disrupted based on the past experiences recorded in its vast knowledge repositories.

Examples of scenarios the AI has been designed to handle include the case of a farming community living on the margins of extreme poverty in a rural area that faced a drought, followed by flooding which damaged crops and killed livestock. Without access to low-interest debt, members of this community would have found recourse in risky migration to urban communities to work in exploitative labor conditions. Another community, fleeing civil conflict, moved to an IDP camp, where they cannot work. Under such circumstances, families might be led to arrange for their underage daughters to enter into marriages to assure their safety in their precarious circumstance. Alex's AI system can predict circumstances in which fragile communities or individuals might become vulnerable to exploitative labor or child marriage, and identify these situations sufficiently far in advance to authorities who are able to prevent these outcomes, whether via financial services, safer employment, or educational interventions. The AI integrates data about weather, migration, community access to credit, civil conflict spread and a myriad of other important factors, and combines this data with survivor-provided insight on moments that can become tipping points into modern slavery. It maps these communities in advance of their reaching a tipping point and engages with authorities to develop effective and context-sensitive interventions.

The AI system developed by Alex is not only able to identify potential trafficking activities and provide recommendations for solutions, it can explain its reasoning backed up by compelling evidence when questioned. It understands that the world is in a constant state of flux, and that the events of the past are not always good predictors of the future when certain important conditions have changed. The AI knows that some data is more trustworthy than other data based on its source, especially in an application like this; while it is designed to use all of the data, not all of it receives equal weight. The system also knows what it does not know and can converse intelligently with experts regarding its points of uncertainty so that they are not misled. There is no such thing as perfect, complete knowledge when fighting a crime like human trafficking.

## Vignette #3 (Collecting and Sharing Heterogeneous Data)

This vignette shows how AI systems handle big data from widely varying sources while keeping in mind the privacy of their subjects in order to enhance quality of life. It is written with a specific focus on the "Societal Drivers for Meaningful Interaction with AI Systems," "Integrated Intelligence," "Knowledge Repositories and Open AI Platforms," and "AI Ethics and Policy." Note how this system is fully context-aware, and can extract substantial commonsense knowledge from all the various sources. It uses provably secure privacy-preserving techniques whenever working with sensitive survivor data.

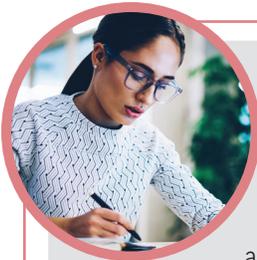

### Vignette 3

Jillian coordinates between an NGO in her home state and a Global Observatory dedicated to identifying, measuring, tracking, and eradicating human trafficking. Their standard practice over a number of years is to capture and integrate a broad range of open and confidential data using an advanced AI system which will prove invaluable in identifying the likely sources of current problems around the world, including Jillian's state, as well as interventions to address them most effectively. This data collected "in the wild" is supplemented by efforts specifically aimed at building substantial parallel datasets in other controlled settings, including AI-ready Hospitals.

The Global Observatory fields AI that is capable of integrating all manner of heterogeneous information at massive scales from around the world, including crime data, police and street outreach observation data, GIS information from community reports, addresses where known victims are living, and known regional prostitution hot spots. It also integrates stories from the news and social media, knowledge of local economic and employment activity, patterns of drug use and abuse and other associated community health data, and events that bring large numbers of outsiders to town. Photos posted on social media and in news reports, and video feeds captured from public spaces are also monitored as needed. The AI system is designed to focus all of this valuable information to the task at hand, is fully context-aware, and can extract substantial commonsense knowledge from all of these sources. It employs causal models it has developed over time that are both broad and deep, covering the reasoning of both an average observer as well as experts in the crime of trafficking. Much of the knowledge in the system was created collaboratively, at both the local and global level, with the AI working side-by-side with law enforcement personal and social service providers from Jillian's NGO and others like it. Other knowledge was added through the use of transfer learning, adapting valuable information from other locales and applications to the problem of identifying and intervening in human trafficking activities in different parts of the world. Because many communities impacted by trafficking are relatively small, special techniques have been developed for machine learning from small amounts of data, too.

The privacy, policy, and legal issues associated with the collection and use of the data have been discussed in public forums where experts and community members have been fully informed and provided their input and, when required, affirmed their own personal approval for use of their data. These usage parameters, laws, and norms are encoded as "meta-data" in the Global Observatory's system, too, and the AI includes them in its reasoning so that it can proactively identify potential violations that place the rights of individuals at risk. The AI also generates companion documentation that tracks method, purpose, language, and security standards over the lifetime of the data. This documentation is stored and regularly updated so that the AI can provide assurances by auditing and tracking who has accessed the data and updated it. It also checks in with data subjects annually to remind them how their data is being used and seek their continued approval. Because interactions with the AI take place in understandable, conversational language, Jillian's NGO has developed trust in the Global Observatory which is reflected in its willingness to share its own data. This trust is further enhanced because the AI uses provably secure privacy-preserving techniques whenever working with data that is sensitive.

## Vignette #4 (Facilitating Ally Networks and Disrupting Criminal Networks)

The next vignette focuses on the need to develop advanced AI techniques that can understand and successfully model the power of social networks, whether good or bad. While researchers generally look for ways AI can improve the functioning of such networks, in the case of criminal networks used to enable human trafficking, the goal is the opposite: to use AI to effectively disrupt these networks. This vignette is written with a specific focus on "Meaningful Interaction with AI Systems," "Self-Aware Learning," "Integrated Intelligence," and "Knowledge Repositories and Open AI Platforms." Note how the AI system can propose a coordinated course of action because of its deep understanding of the traffickers' network, and how it also supports effective collaborations between those working in law enforcement and social services.

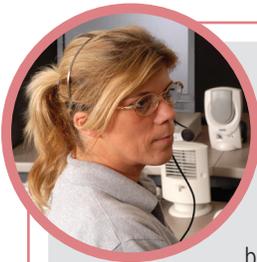

### Vignette 4

Casandra coordinates between law enforcement and a number of local social services organizations in the Greenville community. They have noticed a recent uptick in prostitution reports, but it is not immediately clear what is happening and how to best address the problem. The interactions between Cassandra and her collaborators are in themselves an example of a social network, only in a professional environment and with a serious task at hand. An advanced AI facilitates these interactions, helping each of the participants to identify facts from opinions, monitoring the formation of opinions most useful in this context, and, when called upon, recommending other collaborators and outside resources that will be useful in identifying and evaluating solutions. As a result, the geographically distributed team's communications are much more effective and efficient, and leave far less room for misunderstanding or error.

Fortunately, their standard practice over a number of years is to capture and integrate a broad range of open and confidential data using an advanced AI system which will prove invaluable in identifying the likely source of the current problem, as well as interventions to address it most effectively. Aimed outward, the same AI techniques that facilitate Casandra's professional collaborations are also capable of linking activity on public social networks to external phenomena, such as economics, politics, and crisis events. This information is invaluable in helping to paint a complete picture for Cassandra and her colleagues.

Because the AI system understands collaboration and interaction in social networks, it also understands their use in spreading influence as well as disinformation. The system is able to capture the criminal social network, despite their best efforts to disguise their activities by "getting lost in the noise." While advanced AIs are usually thought of as helping to facilitate interactions between people, when aimed at the traffickers and their customers and victims, this AI is trained for the opposite effect: interfering with those interactions and effectively disrupting the trafficking activity.

Casandra queries the system to help her make sense of what she has been seeing over the past several weeks. The AI is able to provide several plausible explanations, and when pressed further by Casandra can explain its reasoning in ways that are both understandable and will also satisfy the legal requirements for initiating certain actions by law enforcement. The AI system recognizes that local conditions have changed over the past year, and scenarios that might have made sense last year can no longer explain what is happening now. When Casandra needs to dig deeper to determine whether certain key information is trustworthy, the AI system is able to provide the provenance of the data in question, as well as the impact it has had on the recommendations it offered.

**Vignette 4 Continued**

Feeling confident, Casandra recommends a coordinated course of action proposed by the AI to both law enforcement as well as the social services organizations. The goal is to intervene in the trafficking activity to end it, not just cause it to disperse to another location within the municipality, an undesirable outcome the system has already modeled based on past experiences. Because the AI has a deep understanding of the traffickers' network, the disruption is devastatingly effective. As a result, the positive impact of the intervention is maximized and the team working to stop the trafficking activity remain fully informed regarding all aspects of the minute-by-minute progress. The AI system is continually learning, ingesting new data on this latest effort so that it can incorporate feedback as the perpetrators try to adapt their methods; even though they have access to the best consumer-grade computing technologies, their tools are no match for the AI system Casandra and her colleagues have developed.

## Vignette #5 (Privacy, Security, and Ethics)

The following vignette highlights the use of privacy-enhancing technologies (PETs) in AI systems. It is written with a specific focus on "Societal Drivers for Meaningful Interaction with AI Systems," "Self-Aware Learning," "Integrated Intelligence," "Knowledge Repositories and Open AI Platforms" and "AI Ethics and Policy." Note how the AI is also able to provide plain-language explanations of highly technical concepts so Leon can incorporate effective de-biasing techniques, and other PETs to ensure that vulnerable groups are not placed at risk as a result of race, ethnicity, and immigration status.

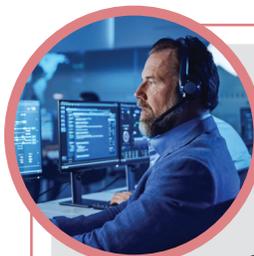

**Vignette 5**

Leon is a specialist working as a team member across several of the organizations appearing in previous vignettes. His responsibility is helping to maintain the long-term privacy, security, and ethical standards surrounding the collection and use of survivor data. Much of the information connected to the fight against human trafficking is extremely sensitive, and the affected individuals have lived portions of their lives — often substantial portions — in positions of great vulnerability. The leakage of any data, whether a small amount of personal data associated with a single survivor, or the knowledge learned from a large group of survivors, could be devastating. The experts in law enforcement and social services know this and respect this, and to gain their trust any AI they work with must share the same understanding. Leon and his colleagues have engineered their AI to meet these expectations.

A fundamental value is the balance between upholding individual privacy rights versus the benefits obtained from the open sharing of individual and group data. In order to come to a uniform consensus, Leon has brought together different stakeholders (companies, NGOs, survivor advocacy orgs, governments, law enforcement, etc.) who worked to develop a shared understanding of what "data" means, especially when the data in question belongs to those in vulnerable groups. Attaching data to individual autonomy is important as this ensures the focus remains on preserving privacy. The group led by Leon has also designed human-centered, AI-powered audit checks to ensure that organizations are responsible in their sharing, management, and (when applicable) discarding of personal data. Throughout there is an emphasis on informed and active consent through ongoing formal mechanisms for data owners to be informed of the ways in which their data is being utilized by different agencies, and for the option to opt-out (or decline to re-opt-in) at any point in time, as they see fit.

> **Vignette 5 Continued**
>
> Since it is now very well understood how AI can incorporate insidious biases that are inherent in data collected across history, Leon has implemented effective de-biasing techniques to ensure that particular groups are not harmed based on race, ethnicity, immigration status and other vital criteria. When potentially damaging biases are discovered in new data to be brought into the system, the AI alerts Leon. This is explanation-based AI, and Leon is able to query, probe, and challenge the system as needed to get to the bottom of the situation, and ensure that ethics is baked into the AI algorithms both at the modular and systems level.
>
> When it comes to pursuing opportunities for beneficial data sharing, Leon, working with the AI, confirms that the appropriate privacy-enhancing technologies (PETs) are in place. Potential options include the use of synthetic datasets, differential privacy, secure multi-party computation, and homomorphic encryption, all of which have been developed on a firm theoretical foundation and stress-tested in the field. The AI is also able to provide plain-language explanations of these highly technical concepts so that average users can understand what the AI is doing and why.
>
> Above all, the metrics, measures, and risk assessments used to evaluate the AI are people-centered, infused with a strong shared sense of human rights from the outset.

## Linkages to the AI Roadmap

We now describe some of the significant AI research challenges and their connections to the visions illustrated in the previous vignettes.

### Modeling and Communicating Mental States

In the context of our vignettes, an AI system needs to work collaboratively with Susan and Casandra, as well as with AI-based teammates (if some of the services it wants to access to assist Susan and Casandra are mediated by their own intelligent agents). It must be fluent in verbal and nonverbal communications, be prepared to engage in mixed-initiative interactions, and be aware of shared mental models. It should be able to predict actions and emotions to the same extent a human expert with knowledge of the situation would.

It takes significant experience for even a professional to be able to discern the mental state of someone they are communicating with. Getting it wrong when attempting to help an individual like Susan who is in a fragile condition, and who has already been manipulated by others, could have tragic consequences. AI is not yet capable of this level of sophistication; even detecting sarcasm is challenging for today's language understanding systems. To accomplish this goal, it will be necessary to progress well beyond the simple text chat interfaces that are now becoming common in online customer service applications. Voice analysis, facial expression and gesture analysis, and knowledge of past interactions and a user's personal history will be valuable components in building intelligent agents that can model and communicate with an awareness of mental states.

One question that might be used to guide this research is "How can AI use a trauma-informed approach to support survivors in such contexts?" Questions like this provide a framework under which both survivors and survivor support organizations could provide inputs to deepen AI's capacity to "understand" how to provide effective support in the most challenging of interactions.

### Alignment with Human Values and Social Norms

To help someone like Susan, an AI system must be able to reason about the ethical implications of the actions it engages in or proposes for her to do. It must also possess an operative understanding of social norms. The AI should take these considerations into account in its planning, and it must also be attuned to unethical behavior it observes on the part of humans and other intelligent agents. Given the skill that traffickers have developed for manipulating people, this will undoubtedly form a major challenge for AI systems.

While AI awareness of ethics, such as in Leon's vignette, needs to permeate all aspects of this research roadmap, this is one area where it comes into primary focus. There are certain values that are universal, others that are country- or culture-specific, and still others that are particular to the individual. For example, when attempting to help someone who has lost her job, it may be appropriate to suggest an open position at a local hamburger restaurant, but not if the user in question is known to be vegetarian. Many existing AIs that serve as intermediaries are designed to optimize the value to the entity on one side of a transaction — usually a business that is selling something — but minimize consideration of the impact to the entity on the other side of the transaction, often the end user. Future AIs must be developed to do a much better job balancing these two considerations, which often trade off against one another.

### Making AI Systems Trustworthy

Trust is an important component in any interaction, but particularly in a situation like Susan's. The intelligent agent helping her must be able to reason about her levels of trust, as well as the trust of other humans it engages to assist her. This involves knowing something about the histories of all involved, especially in similar contexts. It would be natural for Susan to trust some aspects of the AI system and not others, especially if she is encountering new behavior by the AI under stressful circumstances.

An intelligent agent must be trusted in order to be helpful, but the level of trust it receives should be appropriate (not too much, not too little), and in most cases trust must be earned over time as a result of successful interactions with the user such in Casandra's vignette. As a necessary component of building trust, the guidance provided by the AI must be immediately and transparently understandable by the user, and it must appear rational, regardless of the user's level of technical expertise. Mistakes made must be appropriately acknowledged and the AI must learn from them so that similar mistakes are not repeated in the future. The AI must build a level of trust that justifies a vulnerable individual sharing highly confidential information, in a world where it is now well known that many other fielded AIs are being used to exploit personal data for commercial gain and to manipulate public opinion.

### Preventing Undesirable Manipulation

To influence someone like Susan or Leon to take an action, even if it is clearly for their own good, requires persuasion. To build trust, an AI system should use persuasion when it is ethically appropriate. As Susan's agent, the AI should also strive to protect her from intentional manipulation by other humans or AIs.

Nearly all forms of "conversational" interactions with AI can be viewed as a form of manipulation. Even the order in which Internet search results are returned, or music or movie choices are recommended to a consumer, are attempts to influence user behavior. Undesirable manipulations are ones which put the user in more danger, or are chosen because they provide maximal benefits to the party providing the AI even though they are less beneficial to the user. To the extent that the AI is used to monitor human-to-human interactions, e.g., when it has put a user in touch with others who claim to be providing assistance, for example access to housing, or mental health services, or a connection to a support group, then the AI should also be capable of determining whether the claimed benefits to the user are in fact being accrued.

### Supporting Interactions Between People

Beyond the pursuit of AI research to directly address the needs of vulnerable individuals like Susan, some of the most important functions of such AIs will be in connecting users with well-trained human experts who are qualified to assist them. While many of today's most successful smartphone apps are designed with human-to-human communications at their core, they are not architected to take into account the nature of the interactions between individuals, other than perhaps whether someone signals that they "like" what they have seen or read. As a result of research, we expect that AI will develop the capability to more carefully monitor these communications, and to intercede when necessary, either to support discussions that are heading in a positive direction, or to redirect those that are heading in a negative direction as in Casandra's vignette. In doing so, the AI will have to recognize and model facts and goals, including what the individuals are planning and the steps they are taking to implement those plans. This is AI serving as a facilitator.

## Contextualized AI

This topic is related to the earlier discussion on modeling and communicating mental states, however taking into account more of the broader context in which an interaction is taking place. Some attitudes or actions that would appear dangerous or destructive in certain contexts might be perfectly acceptable in others. The AI must know that Susan considers herself at risk, the triggers that might spark a certain extreme reaction, and the steps to take to try to mitigate the situation. Quickly and accurately recognizing such situations, and the range of reactions that would be appropriate, will be vital in the AI's ability to understand and support Susan.

## Explanation and Interpretability

For sophisticated applications in complicated domains like the ones Alex is addressing, an AI system must be prepared to explain its beliefs about the world, and its reasoning in specific cases. It must be open to challenges and prodding by human experts who are justifiably skeptical, or who need a very high degree of certainty before embarking on a dangerous and/or expensive course of action. The users of Alex's intelligent agent must be able to trust the system appropriately and understand its limitations.

In this case, the AI pulls large streams of data from anti-slavery research that indicate where prevalence has increased and under what circumstances (in response to crop failures, civil conflict, displacement, and so forth). With its access to vast knowledge repositories and its advanced machine learning and interaction skills, it can converse with users regarding the respective likelihood of different economic, social, or political crises to create vulnerabilities that increase slavery, trafficking, child labor, or forced marriage.

## Durable Machine Learning Systems

Collecting the data needed to train current machine learning approaches is costly and time consuming. Once the data is in place for a particular problem, there is a natural tendency to use it again and again. The real world is constantly changing, however, and recognizing when the data a system was trained on is no longer representative is a key capability in future AI systems such as the one Alex has built. This is known as "data drift." Alex's system finds ways to exploit broader memory to counter this effect; knowledge that it or others have previously acquired. It is able to adapt from problems it knows how to solve to problems that are new, through concepts like transfer learning. It is able to learn effectively when confronted by new and novel situations it has not encountered before (e.g., a once-in-a-lifetime natural disaster in a certain part of the world that places new stresses on the human population living there). This is what is meant by "durable" machine learning systems.

Referring back to the vignette, the AI system built by Alex and their colleagues is able to track complex interplays of weather patterns (to identify likely spots for crop failures, displacement, and conflict) along with government responses (to identify the likelihood that national or local governments will be able to protect communities in the case of a short- or long-term crisis) and actions by populations (say, through their movements or their social media activities) so that it can flag potential problems before they arise and propose effective ways for ameliorating them.

## Trustworthy Learning / Data Provenance

Alex's AI system leverages decades of knowledge collected from an enormous variety of sources of varying qualities and trustworthiness. It would not be feasible or desirable to attempt to eliminate the data that was not 100% reliable, a notion that itself is even hard to define and variable over time. Instead, future AIs will employ repositories with strong provenance information. AI systems will employ this provenance to improve resilience and support updates and extensions to provenance standards.

The intelligent agent we describe our vignette is able to make reliable predictions and trustworthy recommendations regarding communities vulnerable to exploitation precisely because it is built to understand that not all data is equally good. It is able to communicate with its users the sources of the evidence it is using, and the potential problems that might be present based on the provenance of the data. As additional data becomes available on points of vulnerability, effective interventions, and survivor experiences, these can be used to strengthen the AI's understanding, resolve points of doubt, and better guide the system's users.

### Quantification of Uncertainty

Future AI systems, such as the one Alex has created, will be capable of building and maintaining models of their own competence and uncertainty. It is important to recognize what you know, and what you do not know. This includes uncertainties about the problem formulation and model choice, as well as uncertainties resulting from imperfect and constantly changing data, and uncertainties arising from the unpredictable behavior of other entities operating in the system, human and AI. Because of its sensitive application area, Alex's AI is designed to exhibit a high degree of robustness to such uncertainties, and is open and honest with its users about the gaps in its knowledge and the limits of its recommendations.

In this case, the intelligent agent created by Alex and their colleagues is able to build uncertainty into models based on human behavior (political interventions, public interest in supporting international relief efforts). It can also quantify the uncertainty of long-term versus short-term outcomes of interventions; in other words, while a community may be supported through one crisis, most vulnerable communities experience continual and recurring crises.

### The Science of Integrated Intelligence

Many of the AI systems in these vignettes have power because of their access to huge, rich collections of widely varying data on human activities and other heterogeneous information recorded in AI-ready repositories. Any data or experience on which a social services or law enforcement expert might base a decision is included in these repositories. Collecting and storing such data is relatively easy. Structuring it for effective use and protecting it from potential abuses is an enormous research challenge.

There are myriad past experiences that inform how an individual reacts and what they need to do to make successful progress toward a goal when confronted by the day-to-day decisions and challenges in their life. Some of these experiences are acquired over a lifetime, while others are much more recent. They span a range of broad categories, including health concerns (of the individual, their family and friends, their community), personal security, food security, housing, finances and job prospects, education, etc. All of these are clearly relevant to the situation Susan finds herself in; some parts of the solution to her current problem may very well lie in her past. Today's systems, often embodied in smartphone apps, capture and record narrow slices of relevant data, most often for the purposes of marketing a product to customers. But they do not maintain this information for long, nor do they do attempt to integrate it across a variety of sources since this is not the driving economic force. An advanced AI that can help these people in these vignettes must be able to construct and employ repositories of useful information and experiences that are orders of magnitude broader and deeper than what is possible today. These repositories must be designed in ways that preserve her personal privacy while still being effective under the most difficult and stressful of circumstances.

### Knowledge Capture and Dissemination

The intelligent agent we presented earlier depends on purpose-built knowledge repositories for its target domain, the fight to end human trafficking. With a specific domain in mind, it is possible to identify the sources needed to create such knowledge bases, and then extract entities and the relations between them yielding very large knowledge graphs which provide an important tool and a valuable source for insights. It should not be surprising that AI systems not only use these resources but they can also assist in assembling and structuring knowledge repositories and knowledge graphs and help to validate and maintain the knowledge they represent on an ongoing basis. Alex's AI also has the ability to extract significant amounts of commonsense knowledge from text, images, videos, and other unstructured formats, vital skills in attempting to identify indicators of trafficking and support efforts to end it.

In the case of our vignette, the AI captures knowledge from weather patterns, including predictive analysis of climate change and rising sea levels. Data regarding historical farming productivity and the seasonal supply/demand for food is also maintained. In addition, it is useful to capture patterns of communication that indicate community preparation to enter into risky labor, such as SMS conversations with prospective employers or signs of other common recruiting activities based on prior experiences contained in the knowledge base. Connections between individuals known to be engaged in trafficking or on the fringe of such activities can be discerned through the entity-relationship knowledge graph, and this information can also be used in making predictions and suggesting corrective actions.

### Heterogeneous Knowledge

The AI system Alex and their colleagues built parallels other advanced AIs employing causal models that are broad and deep enough for use across diverse application areas. All are capable of integrating vast stores of heterogeneous knowledge. For example, similar capabilities would be required for intelligent agents working in education, medicine, eldercare, and design. Repositories of causal models that cover both everyday phenomena and the performance of experts enable Alex's intelligent agent to act as a collaborator with its human partners.

As affected communities and support organizations feed data into the system on interventions offered and their impact on reducing vulnerability, the AI is able to learn from their experiences. It harvests the most effective interventions and offers insights regarding their efficacy to support future investigations, and also records efforts that fell short of expectations with the goal of not repeating the same mistakes twice. This information will arrive in a wide range of heterogeneous formats which must be processed for inclusion in the knowledge repository.

### AI-Ready Data Repositories

Several times in this section we have referenced "AI-ready" data repositories, which is another recommendation drawn from the AI Roadmap [AI19]. These are a vital component of the shared research infrastructure that are key enablers. They include open datasets that address real-world problems relevant to the matter at hand; all of the subtasks we have described previously in the context of the vignette are supported by associated AI-ready repositories. Producing such repositories is a significant effort in itself. Some of the challenges include managing sharing for domains and tasks that have complex and sometimes sensitive data, such as social communications, interacting behaviors, and real-world goal accomplishment. While a large component of what is needed here is infrastructure-building, there are also deep research questions in privacy and security given the extremely sensitive nature of the data associated with survivors and victims of trafficking [see, e.g., BD17, MDC20, NPR15].

### AI Ethics and Policy

In both Jillian and Leon's vignettes, the importance of incorporating ethics and related responsibility principles as central elements in the design and operation of AI systems are highlighted. AI ethics is concerned with responsible uses of AI, incorporating human values, respecting privacy, universal access to AI technologies, addressing AI systems bias particularly for marginalized groups, the role of AI in profiling and behavior prediction, as well as algorithmic fairness, accountability, and transparency.

AI privacy, which Leon is focused on, is concerned with the collection and management of sensitive data about individuals and organizations, the use and protection of sensitive data in AI applications, and the use of AI for identification and tracking of people. Designers should give considerations of societal (ethical and political) values the same prominence as traditional drivers such as performance, accuracy, and scalability. By doing so we may advance generations of technologies that are less likely to cause harm and increase privacy for survivors [IA20].

### Life-long and Life-Wide Personal Assistant Platforms

An exciting application area for AI with potential for high impact in society are personal assistants (LPAs). LPAs are AI systems that assist people in their daily lives at work and home, for leisure and for learning, for personal tasks and for social interactions. Their overall design is also a fundamental research challenge, since an LPA could be a single AI system (like Susan's AI) with a diverse range of capabilities or a collection of independently developed specialized systems (like Jillian or Casandra's) with coordination mechanisms for sharing user history and personalization knowledge. Open platforms for continuous data collection and dynamic experimentation would significantly accelerate research, not only in this important application of AI but also in basic AI research.

### Societal Drivers for AI Systems

There is substantial discussion of enhanced health and quality of life as key societal drivers for research in the AI Roadmap [AI19]. The application envisioned there is from healthcare, which is also applicable in the case of trafficking victims. AI systems should have an understanding of a survivor's health issues and the risks they face in order to provide the best assistance, just as in Susan's vignette. There are also clear parallels between health and trafficking when it comes to data that is relevant and AI techniques to process it, such as in Jillian's vignette, where what is learned in the former can be useful in the latter — e.g., many of the same collection, storage, security, and privacy issues will arise in both applications.

## Additional Considerations

While the previous discussion focused on visions of the future where AI research can make significant contributions in the fight against modern slavery, many practical considerations are important to keep in mind on the path toward that goal.

There are different interventional approaches to addressing even one type of modern slavery. Evaluation is hard because these networks are fast moving, fast changing, and adaptable. Baseline measures are often lacking, control groups are limited and are difficult to set up, and the ability to monitor the sustained success of the intervention over time and/or whether it has just displaced the trafficking activity is often impossible [EP07]. This is despite studies highlighting that such displacement does in fact occur. A prominent example is the Cardiff Violence Prevention Model, which provides a way for communities to gain a clearer picture about where violence is occurring by combining and mapping both hospital and police data on violence. "But more than just an approach to map and understand violence, the Cardiff Model provides a straightforward framework for hospitals, law enforcement agencies, public health agencies, community groups, and others interested in violence prevention to work together and develop collaborative violence prevention strategies [CDC2020]."

As part of their standard practice, user-centered design research projects growing out of these efforts should hire survivors as compensated project developers and participants. These survivors may be residing in shelters or otherwise connected to service organizations. See, e.g., [SA20]. There should be a commitment to using lived experience testimony from survivor-consultants and applying it to large datasets in an effort to reduce points of contact and end traffickers' ability to exploit vulnerabilities.

The lived experiences of survivors provide raw data, which is critical, and their opinions about what constitutes effective research and interventions will prove invaluable. Computing researchers have experience incorporating end users in their work, but inclusion of survivors will be fundamentally different. This is a matter that deserves its own serious considerations. An example of one approach for engaging with the survivor community would be to:

1. Generate survivor-informed frameworks for choosing which open field self-descriptors (e.g., cultural, personal, demographic) should be presented to the global research group to establish survivor identity profiles as part of a data collection activity.

2.  Design paid, opt-in engagement for a wider audience of survivors globally, to provide their self-descriptors and to provide details about their recruitment and exploitation to establish patterns of perpetrator behavior.

3.  Implement paid, opt-in engagement for a wider audience of survivors globally to define the "correct" answer or the "need to hear / want to hear" answer for training machine learning algorithms to be survivor-informed. This data is gathered alongside the specific self-identified contextual data points to enable future complex analysis. The AI developed in these exercises could be applied across many different use cases to support survivor well-being and prevent / fight exploitation, as well as identify areas where investment and resources will make the most impact.

4.  Combine the outcomes of Steps 1-3 to identify lookalike circumstances of perpetrator actions (e.g., offering a temporary offsite work in a community that has experienced a major natural disaster) and/or self-descriptor categories likely to be targeted by those actions (people living below the poverty line in non-majority ethnic groups), and develop improved warning / awareness / communication mechanisms to reach the vulnerable population at a high risk of that type of exploitation.

Another set of considerations is the importance of international collaboration. Human trafficking is a crime that pays no heed to national borders. Solutions can only be achieved through international collaborations at all levels in the fight: between law enforcement agencies, NGOs, social service organizations, researchers, and funding agencies.

## Call to Action

Modern slavery is one of the most persistent and vexing problems facing humanity. AI will not be a panacea, but it can be a valuable tool in the fight. With the tremendous investments now being made in AI and AI-related research, we see significant value in instilling a greater awareness of the open challenges and potential applications among those performing such research and those helping to guide it. Likewise, practitioners and other stakeholders should be aware of what might be possible with AI in a few years, and hopefully be better positioned to help contribute to realizing these visions.

Communication and collaboration are key, and these are among the primary goals of Code 8.7 [Code8.7]. For those working on the computing side of the equation, our closing advice for those who wish to contribute is to develop a collaborative research plan designed to address human trafficking using AI, one which keeps central the perspective of survivors and considers how outputs can be made available for implementation in real-world contexts. It is also vital that developers of new communications and data collection technologies are aware of the human rights, privacy, and ethical considerations of their use, including the potential of new technologies to facilitate human trafficking — or prevent it.

## Acknowledgements

The organizers would like to thank James Cockayne (Nottingham) and the workshop participants for their input in this report.

CCC / CODE 8.7## Workshop Participants

| First Name | Last Name | Email | Affiliation |
|---|---|---|---|
| Elisa | Bertino | bertino@purdue.edu | Purdue University |
| Nadya | Bliss | Nadya.Bliss@asu.edu | ASU |
| Jessie | Brunner | jbrunner@stanford.edu | Stanford |
| Sherrie | Caltagirone | sherrie@globalemancipation.ngo | Global Emancipation Network |
| James | Cockayne | cockayne@unu.edu | United Nations University |
| Sandra | Corbett | scorbett@cra.org | CRA/CCC |
| Minh | Dang | minh@survivoralliance.org. | Survivor Alliance |
| Julia | Deeb-Swihart | jdeeb3@gatech.edu | Georgia Tech |
| Khari | Douglas | kdouglas@cra.org | CRA/CCC |
| Artur | Dubrawski | awd@andrew.cmu.edu | CMU |
| Alice | Eckstein | eckstein@unu.edu | United Nations University |
| Alex | Endert | endert@gatech.edu | Georgia Tech |
| James | Goulding | James.Goulding@nottingham.ac.uk | University of Nottingham |
| Lorenzo | Guarcello | guarcello@ilo.org | International Labour Organization |
| Peter | Harsha | harsha@cra.org | CRA/CCC |
| Kevin | Hong | Hong.Kevin.C@dol.gov | Labor |
| Edward | Huang | chuang10@gmu.edu | GMU |
| Jessica | Hubley | jessica@anniecannons.com | Annie Cannons |
| Julia | Jeng | cjeng2@gmu.edu | GMU |
| Ran | Ji | rji2@gmu.edu | GMU |
| Matt | Kammer-Kerwick | mattkk@ic2.utexas.edu | UT at Austin |
| Mayank | Kejriwal | kejriwal@isi.edu | USC |
| Georgia Ann | Klutke | gaklutke@nsf.gov | NSF |
| Fred | Kronz | fkronz@nsf.gov | NSF |
| Dan | Lopresti | lopresti@cse.lehigh.edu | Lehigh |
| Kayse | Maass | k.maass@northeastern.edu | Northeastern University |
| Ross | Maciejewski | rmacieje@asu.edu | ASU |
| Anjali | Mazumder | amazumder@turing.ac.uk | Turing |

| Sarah | Mendelson | smendelson@cmu.edu | CMU |
|---|---|---|---|
| Florian | Ostmann | fostmann@turing.ac.uk | Turing |
| Osman | Ozaltin | oyozalti@ncsu.edu | NC State University |
| Michael | Paley | mike@senderoadvisory.com | Sendero Advisory |
| Winifred | Poster | wposter@wustl.edu | Washington University St.Louis |
| Amy | Rahe | amy@survivoralliance.org | Survivor Alliance |
| Dominique | Roe-Sepowitz | Dominique.Roe@asu.edu | ASU |
| Ann | Schwartz | aschwartz@cra.org | CRA/CCC |
| Danielle | Smalls | dsmalls@uncharted.software | Uncharted |
| Gavin | Smith | Gavin.Smith@nottingham.ac.uk | University of Nottingham |
| Shannon | Stewart | shannon@gfems.org | Global Fund to End Modern Slavery |
| Mitali | Thakor | mthakor@wesleyan.edu | Wesleyan |
| Kathleen | Vogel | kvogel2@umd.edu | UMD |
| Helen | Wright | hwright@cra.org | CRA/CCC |
| Shanji | Xiong | Shanji.Xiong@experian.com | Experian |

# NOTES

# NOTES

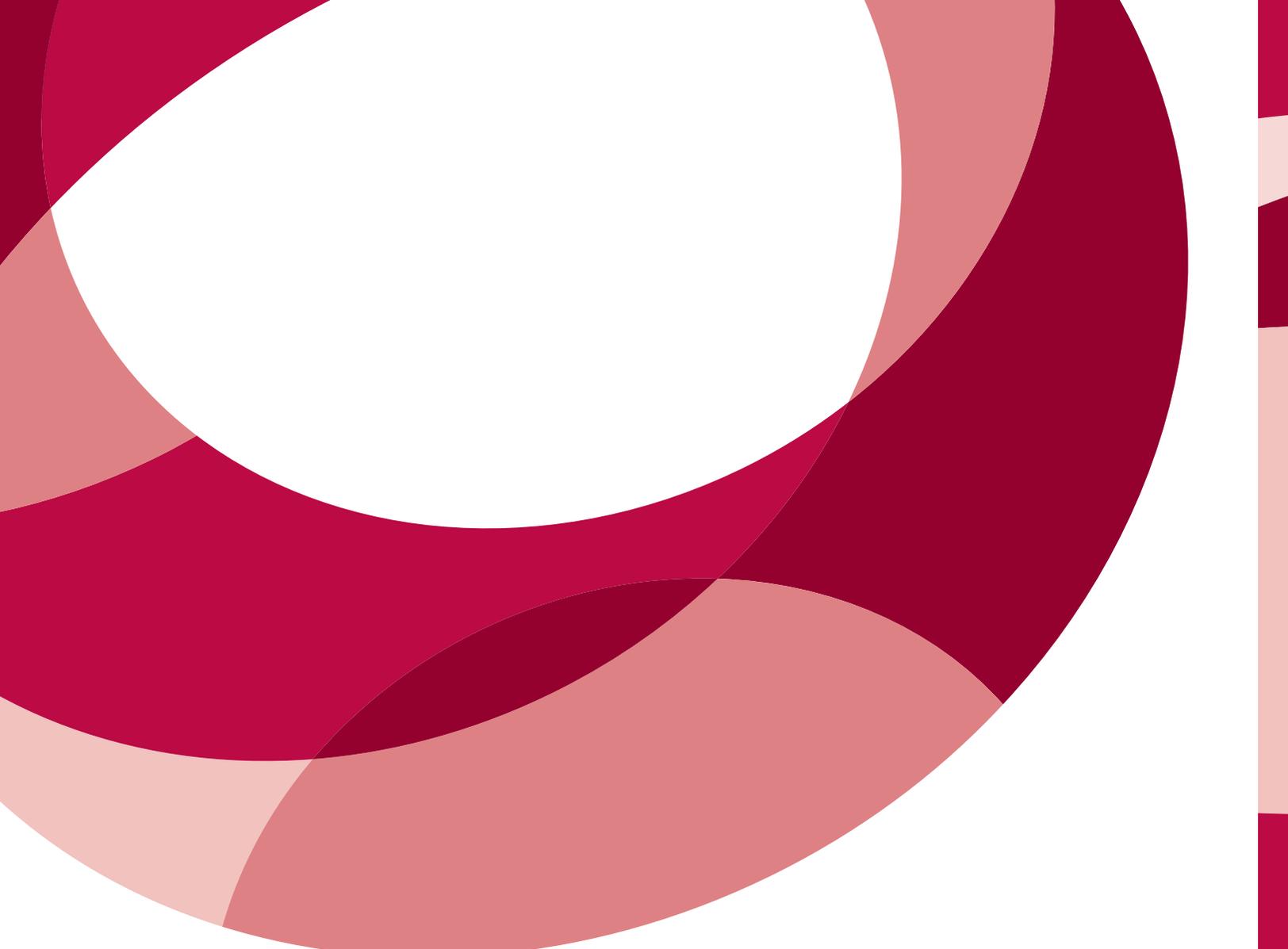

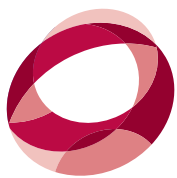

**CCC**
Computing Community Consortium
Catalyst

1828 L Street, NW, Suite 800
Washington, DC 20036
P: 202 234 2111 F: 202 667 1066
www.cra.org cccinfo@cra.org